\documentclass[prd, showpacs,twocolumn,floatfix,nofootinbib]{revtex4}
\usepackage{epsfig}
\usepackage{graphicx}
\usepackage{amsfonts}
\usepackage{mathrsfs}

\begin{document}

\newcommand{\be}{\begin{equation}}
\newcommand{\ee}{\end{equation}}
\newcommand{\bea}{\begin{eqnarray}}
\newcommand{\eea}{\end{eqnarray}}
\newcommand{\nnb}{\nonumber}
\renewcommand{\thefootnote}{\fnsymbol{footnote}}
\def\lsim{\raise0.3ex\hbox{$\;<$\kern-0.75em\raise-1.1ex\hbox{$\sim\;$}}}
\def\gsim{\raise0.3ex\hbox{$\;>$\kern-0.75em\raise-1.1ex\hbox{$\sim\;$}}}
\def\Frac#1#2{\frac{\displaystyle{#1}}{\displaystyle{#2}}}
\def\no{\nonumber\\}
\def\slash#1{\ooalign{\hfil/\hfil\crcr$#1$}}
\def\ep{\eta^{\prime}}
\def\susy{\mbox{\tiny SUSY}}
\def\sm{\mbox{\tiny SM}}
\def\pslash{\rlap{\hspace{0.02cm}/}{p}}
\def\qslash{\rlap{/}{q}}
\def\kslash{\rlap{\hspace{0.02cm}/}{k}}
\def\lslash{\rlap{\hspace{0.011cm}/}{\ell}}
\def\nslash{\rlap{\hspace{0.02cm}/}{n}}
\def\Pslash{\rlap{\hspace{0.065cm}/}{P}}

\textheight      250mm  
\vskip0.5pc

\title{Prediction of $B_c\to D\pi$ in the PQCD approach}
\author{Jian-Feng Cheng\footnote{Email: chengjf@mail.ihep.ac.cn}}
\address{    Institute of High Energy Physics, CAS, P.O. Box 918(4),
Beijing 100049, China}

\author{Dong-Sheng Du  and Cai-Dian L\"u }
 \affiliation{     CCAST (World Laboratory), P.O.
Box 8730, Beijing 100080, China; }
 \affiliation{   Institute of
High Energy Physics, CAS, P.O. Box 918(4), Beijing 100049,
China\footnote{Mailing address.}}

\date{\today}
\begin{abstract}
We investigate the branching ratios and direct CP asymmetries of
$B_c^+\to D^0\pi^+$ and $B_c^+\to D^+\pi^0$ decays in the PQCD
approach. All the diagrams with emission topology or annihilation
topology are calculated strictly. A branching ratio of $10^{-6}$
and $10^{-7}$ for $B_c^+\to D^0\pi^+$ and $B_c^+\to D^+\pi^0$
decay is predicted, respectively. Because of the different weak
phase and strong phase from  penguin operator and two kinds of
tree operator contributions, we predict a possible large direct CP
violation: $A^{\rm dir}_{cp} (B_c^\pm \to D^0 \pi^\pm )\approx
-50\%$ and $A^{\rm dir}_{cp} (B_c^\pm \to D^\pm \pi^0 )\approx
25\%$  when $\gamma=55^\circ$, which can be tested in the coming
LHC.
\end{abstract}

\maketitle
\noindent

\section{Introduction}

The charmless $B$ decays provide a good platform to test the
Standard Model (SM) and study the CP violation, which arouses
physicists' great interest and has been discussed in the
literature widely. But how about the $B_c$ decays, the $b$ quark
of which has similar property with that of $B$ meson? There are
some events of $B_c$ at Tevatron \cite{tevatron} and will be a
great number of events appearing at LHC in the foreseeable future.
The progress of the experiments makes us to think of a question:
what will be the theoretical prediction on the two-body
non-leptonic $B_c$ decays?

Different from $B$ and $B_s$ meson, $B_c$ meson consists of  two
heavy  quarks $b$ and $c$, which can decay individually. Because
of the difference of mass, lifetime and the relative  CKM matrix
element between $b$ and $c$ quark, the decay rate of the two
quarks is different, which determines the unique property of $B_c$
decays. Though $c$ quark's mass is about one third of $b$ quark,
leading to a suppression of $(M_c/M_b)^5$, the decay of $c$~quark
can not be ignored because the corresponding CKM matrix element
$V_{cs}$ is larger than that of $b$ quark: $V_{ub},V_{cb}$.
Because of the small mass of $c$ quark, the decay of $c$ quark is
nearly at non-perturbative scale, where there is great theoretical
difficulty. Now we  study the $b$ quark decay first and leave the
study of $c$ quark decay to the future.

In recent years, a great progress has been made  in studying
two-body non-leptonic $B$ decays in perturbative QCD approach
(PQCD) \cite{pqcd,pqcd-lu}, QCD factorization
 \cite{qcdf} and soft collinear effective theory (SCET)
  \cite{scet}. Though $B_c$ decay has been studied  \cite{du} in Naive
Factorization   \cite{nf,ali} many years ago, no one applies the
method developed recently in such processes. In this paper we will
use $B_c\to D\pi$ as an example to discuss the $B_c$ decays in the
PQCD approach.

The $B_c\to D\pi$ decay provides opportunities to study the direct
CP asymmetry. Different from the $B$ decays   \cite{pqcd-lu},
$B_c\to D\pi$ has the direct CP asymmetry even without considering
the contributions from penguin operators, because the tree
contributions from the annihilation topology provide not only the
strong phase, but also the different weak phase. According to the
power counting rule of PQCD, the tree contributions from the
annihilation topology is   power suppressed. But the  larger CKM
matrix elements $|V_{cb}|$ enhances the contribution to make it
larger than the penguin contributions, so the direct CP asymmetry
of $B_c\to D\pi$ can be very large, which is found in our
numerical analysis.

\begin{figure}
\vskip -4cm
\begin{minipage}[t]{0.43\textwidth}
   \epsfxsize=10cm
   \centerline{\epsffile{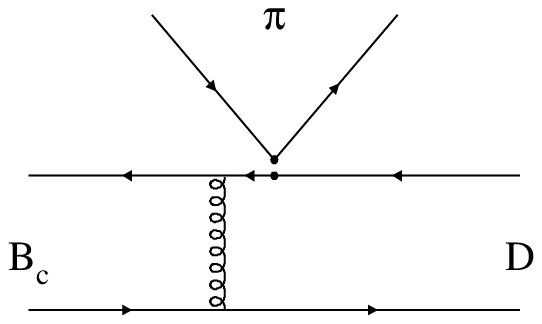}}
   \end{minipage}
\vskip -6cm
\begin{minipage}[t]{0.43\textwidth}
   \epsfxsize=10cm
   \centerline{\epsffile{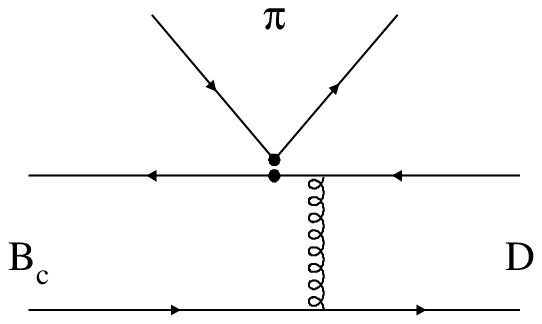}}
   \end{minipage}
   \vspace*{-2.4cm}
\caption[dummy]{\label{fe}Form factor in $B_c\to D\pi$.}
\end{figure}

The study of $B_c$ decay also provides opportunities to test $k_T$
factorization in PQCD approach. According to numerical analysis in
the literature, the form factor contributions from Fig.\ref{fe}
usually dominate the whole decays. In the same way, the form
factor also gives the main contributions in the $B_c\to D \pi$
decay according to our numerical analysis. Since the $B_c$ meson
consists of two heavy quarks, the effect of $k_T$ in the $B_c$
meson can be ignored and the form factor $B_c\to D$ only includes
the $k_T$ contributions
  from $D$ meson. So it is   easier to study
how important the $k_T$ contributions are in $B_c$ decays than
that in the $B$ decays because the latter need to consider both
$k_T$ contributions of $B$ and $D$ meson.

The  $B_c\to D\pi$ decay also provides a good platform to study
the $D$ meson's wave function. The $D$ meson's mass $M_D$ is not
so large that it is hard to get the ideal wave function of $D$
meson by the expansion of $1/M_D$ as in $B$ meson. People use the
form fitted
  from experimental data generally. Such discussion has been done by
Ref.  \cite{li-bd} in the form factor $B\to D$ transition. It is
better to discuss $D$ meson wave function in $B_c\to D \pi $ for
two reasons: one is that the hierarchy between $M_{Bc}$ and $M_D$
($M_{Bc}\gg M_D$) guarantees us to apply the $k_T$ factorization
theorem in this process, the other is that the wave function of
$B_c$ is clean, which eliminates the possible uncertainty from
$B_c$ meson. The experiment of $B_c$ decays will test how
reasonable it is. As the only parameter with large uncertainty,
the wave function of $D$ meson need further theoretical
investigation.

\section{Framework }

\begin{figure}
\vskip -4cm
\begin{minipage}[t]{0.43\textwidth}
   \epsfxsize=10cm
   \centerline{\epsffile{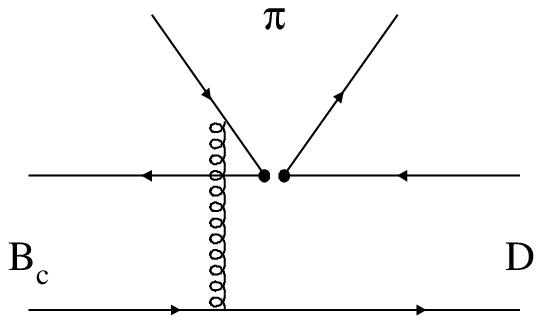}}
   \end{minipage}
\vskip -6cm
\begin{minipage}[t]{0.43\textwidth}
   \epsfxsize=10cm
   \centerline{\epsffile{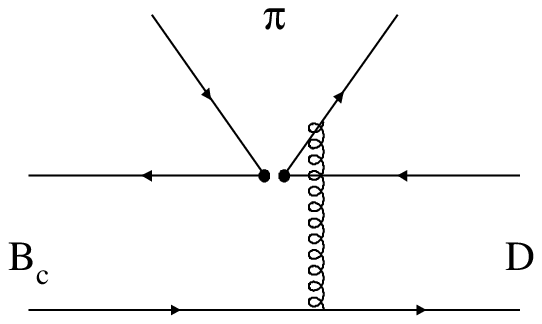}}
   \end{minipage}
   \vspace*{-2.4cm}
\caption[dummy]{\label{nfe}Non-factorizable emission topology
in $B_c\to D\pi$.}
\end{figure}

\begin{figure}
\vskip -3.6cm
\begin{minipage}[t]{0.43\textwidth}
   \epsfxsize=10cm
   \centerline{\epsffile{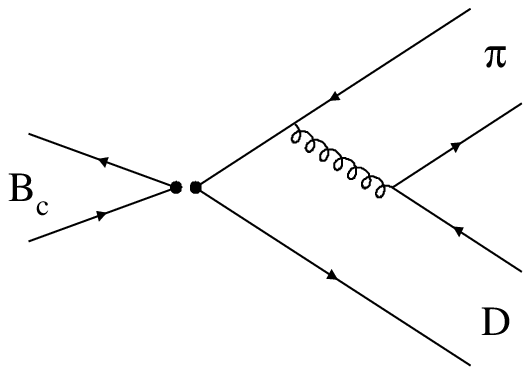}}
   \end{minipage}
\vskip -6cm
\begin{minipage}[t]{0.43\textwidth}
   \epsfxsize=10cm
   \centerline{\epsffile{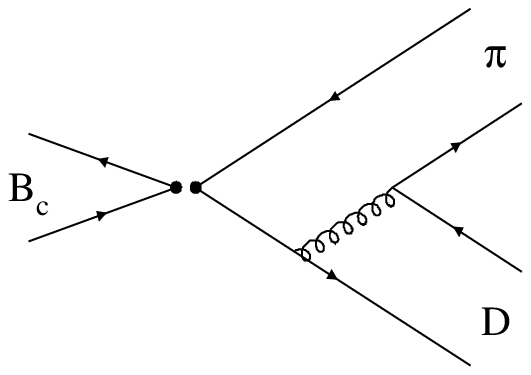}}
   \end{minipage}
   \vspace*{-2.8cm}
\caption[dummy]{\label{fa}Factorizable annihilation topology
in $B_c\to D\pi$.}
\end{figure}

\begin{figure}
\vskip -3.6cm
\begin{minipage}[t]{0.43\textwidth}
   \epsfxsize=10cm
   \centerline{\epsffile{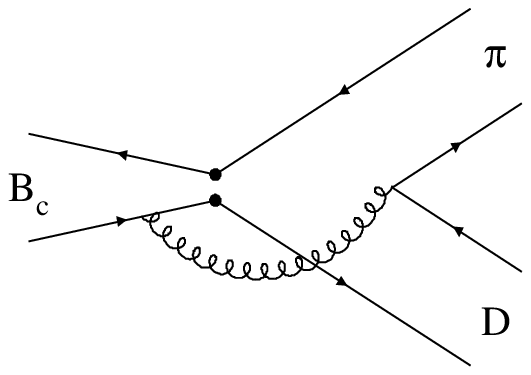}}
   \end{minipage}
\vskip -6cm
\begin{minipage}[t]{0.43\textwidth}
   \epsfxsize=10cm
   \centerline{\epsffile{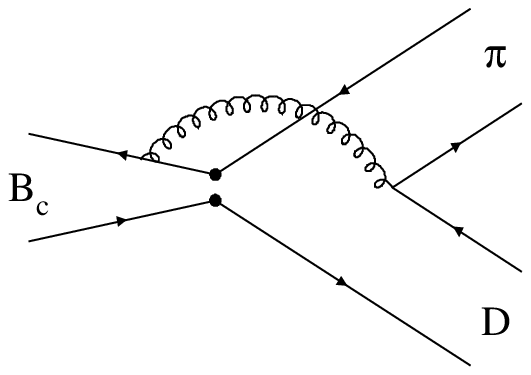}}
   \end{minipage}
   \vspace*{-2.8cm}
\caption[dummy]{\label{nfa}Non-factorizable annihilation topology
in $B_c\to D\pi$.}
\end{figure}

The hard amplitudes of these decays contain factorizable diagrams
(Fig.~\ref{fe}), where hard gluons attach the valence quarks in
the same meson, and non-factorizable diagrams (Fig.~\ref{nfe}),
where hard gluons attach the valence quarks in different mesons.
The annihilation topology is also included, and classified into
factorizable (Fig.~\ref{fa}) and non-factorizable (Fig.~\ref{nfa})
ones according to the above definitions.

In the calculations of all the diagrams, we can ignore the $k_T$
contributions of $B_c$ meson because it consists of two heavy
quarks. Furthermore, we can suppose the two quarks $\bar b$ and
$c$ of $B_c^+$ meson to be on the mass shell approximately and
treat the wave function of $B_c$ meson as $\delta$ function for
simplicity, so we can integrate the wave function $B_c$ out and
the $k_T$ factorization form turns into
\begin{eqnarray}
&&{\rm Form \ factor}\nonumber\\
&&\sim \int d^4 k_1 \, \Phi_D(k_1)\, C(t)\, H(k_1,t), \\
&&{\rm Other \ topology}\nonumber\\
&&\sim \int d^4 k_1 d^4 k_2 \, \Phi_D(k_1)\Phi_\pi(k_2) \, C(t)\, H(k_1,k_2,t) ,
\end{eqnarray}
where $k_{1(2)}$ is momentum of light (anti) quark of $D(\pi)$
meson. The non-factorizable topology includes two kinds of
topology: emission topology (Fig.~\ref{nfe}) and annihilation
topology (Fig.~\ref{nfa}). In the above equations, we sum over all
Dirac structure and color indices. The hard components consist of
the hard part ($H(t)$) and harder dynamics ($C(t)$), the former
$H(t)$ can be calculated perturbatively; the latter $C(t)$ is the
Wilson coefficients which runs from electro-weak scale $M_W$ to
the lower factorization    scale $t$. $\Phi_M$ is the wave
function of $D$ and $\pi$ meson, including the non-perturbative
contributions in the $k_T$ factorization.

Through out the paper, we use the light-cone coordinate to
describe the meson's momentum in the rest frame of the $B_c$
meson. According to the conservation of four-momentum, we get the
momentum of three mesons $B_c$, $D$ and $\pi$ up to the order of
$r_2^2$ ($r_2=M_D/M_{B_c}$)  as following:
\begin{eqnarray}
P_{B_c} &=& {M_{B_c}\over \sqrt{2}}(1,1,\bf{0}_T), \nonumber\\
P_{2} &=& {M_{B_c}\over \sqrt{2}}(1,r_2^2,\bf{0}_T), \nonumber\\
P_{3} &=& {M_{B_c}\over \sqrt{2}}(0,1-r_2^2,\bf{0}_T),
\end{eqnarray}
where we have neglected the small mass of pion and higher order
terms of  $r_2$. Such approximation will be used in the whole
paper.

\section{Calculation of amplitudes}

\subsection{Wave Function}

$B_c$ meson consists of two heavy quarks such that the small
${\bar \Lambda}_{\rm QCD}$ can be ignored (${\bar \Lambda}_{\rm
QCD}=M_{Bc}-M_b-M_c\ll M_c\ \rm or \ M_b$), so as the quark
transverse momentum $k_T$. In principal there are two Lorentz
structures in the B or $B_c$ meson wave function. One should
consider both of them in calculations. However, it can be argued
that one of the contribution   is numerically small
\cite{kurimoto}, thus its contribution can be neglected.
Therefore, we only consider the contribution of one Lorentz
structure, such that we can reduce the number of input parameters
\begin{eqnarray}
\Phi_{B_c}(x) &=& {i\over 4 N_c} (\pslash_{B_c} +M_{B_c}) \,
\gamma_5\,  \delta(x-M_c/M_{B_c}) .
\end{eqnarray}
The other two mesons' wave functions read:
\begin{eqnarray}
 \Phi_{D}(x,b) &=&  \frac{i}{\sqrt{2N_c}}
  \gamma_5 (\not \! P_2+ M_{D})  \phi_{D}(x,b), \\
\Phi_{\pi}(x) &=&  \frac{i}{\sqrt{2N_c}}\big[  \gamma_5 \not \! P_3
\phi_\pi(x) + M_{0\pi} \gamma_5 \phi_\pi^p(x) \nonumber\\
 &&+ M_{0\pi} \gamma_5 (\nslash_-\nslash_+ - 1)\phi_\pi^\sigma(x)
\big] ,
\end{eqnarray}
where $N_c = 3$ is color degree of freedom, and $M_{0\pi} =
M_\pi^2/(m_u + m_d)$, $n_- = (0,1,{\bf 0}_T ) \propto P_3$, $n_+ =
(1,0,{\bf 0}_T)$, $\epsilon^{0123} = 1$.

The momentum fraction of the light quark in the three mesons can
be defined as: $x_1=k_c/P_{B_c}, x_2=k_2^+/P_2^+,
x_3=k_3^-/P_3^-$. In the $B_c$ meson, there are also another
relation between $x_1$ and $r_b=M_b/M_{B_c}$: $x_1+r_b=1$.

\subsection{Effective Hamiltonian}

The effective Hamiltonian for the flavor-changing $b\to d$
transition is given by  \cite{buras}
\begin{eqnarray}
H_{{\rm eff}}&=& {\frac{G_{F}}{\sqrt{2}}}\sum_{q=u,c}V_{q}\bigg[ C_{1}(\mu)
O_{1}^{(q)}(\mu )+C_{2}(\mu )O_{2}^{(q)}(\mu )\nonumber\\
&&+\sum_{i=3}^{10}C_{i}(\mu)
O_{i}(\mu )\bigg] \;,  \label{hbk}
\end{eqnarray}
with the Cabibbo-Kobayashi-Maskawa (CKM) matrix elements
$V_{q}=V_{qd}V_{qb}^{*}$ and the operators
\begin{eqnarray}
&&O_{1}^{(q)}=(\bar{d}_{i}q_{j})_{V-A}(\bar{q}_{j}b_{i})_{V-A}\;,\;\;\;\;\;
\;\;\nonumber\\ &&O_{2}^{(q)}=(\bar{d}_{i}q_{i})_{V-A}(\bar{q}_{j}b_{j})_{V-A}\;,
\nonumber \\
&&O_{3}=(\bar{d}_{i}b_{i})_{V-A}\sum_{q}(\bar{q}_{j}q_{j})_{V-A}\;,\;\;\;
\nonumber\\ &&O_{4}=(\bar{d}_{i}b_{j})_{V-A}\sum_{q}(\bar{q}_{j}q_{i})_{V-A}\;,
\nonumber \\
&&O_{5}=(\bar{d}_{i}b_{i})_{V-A}\sum_{q}(\bar{q}_{j}q_{j})_{V+A}\;,\;\;\;
\nonumber\\ &&O_{6}=(\bar{d}_{i}b_{j})_{V-A}\sum_{q}(\bar{q}_{j}q_{i})_{V+A}\;,
\nonumber \\
&&O_{7}=\frac{3}{2}(\bar{d}_{i}b_{i})_{V-A}\sum_{q}e_{q} (\bar{q}%
_{j}q_{j})_{V+A}\;,\;\nonumber\\ &&O_{8}=\frac{3}{2}(\bar{d}_{i}b_{j})_{V-A}
\sum_{q}e_{q}(\bar{q}_{j}q_{i})_{V+A}\;,  \nonumber \\
&&O_{9}=\frac{3}{2}(\bar{d}_{i}b_{i})_{V-A}\sum_{q}e_{q} (\bar{q}%
_{j}q_{j})_{V-A}\;,\;\nonumber\\ &&O_{10}=\frac{3}{2}(\bar{d}_{i}b_{j})_{V-A}
\sum_{q}e_{q}(\bar{q}_{j}q_{i})_{V-A}\;,
\end{eqnarray}
with $i$ and $j$ being the color indices. Using the unitary
condition, the CKM matrix elements for the penguin operators
$O_{3}$-$O_{10}$ can also be expressed as $V_{u}+V_{c}=-V_{t}$.

The $B_c\to D \pi $ decay rates have the expressions
\begin{equation}
\Gamma =\frac{G_{F}^{2}M_{B_c}^{3}}{32\pi }|A|^{2}\; . \label{dr1}
\end{equation}
 The decay amplitude $A$ of $B_c\to D\pi$  process from all the diagrams
 can be expressed in the following:
\begin{eqnarray}
A_{D^0\pi^+} &=&
V_u (f_\pi F^T_{e1} + M^T_{e1}) +V_c (f_{B_c} F^T_a + M^T_a)
\nonumber\\
&&\hspace{-0.3cm}- V_t \left( f_\pi F^{P1}_{e1} + f_\pi F^{P3}_{e1}
+  M^{P1}_{e1} +  M^{P2}_{e1} \right. \label{d0pi}\\
&&\left.+f_{B_c} F^{P1}_a +f_{B_c} F^{P3}_a+ M^{P1}_a +
M^{P2}_a\right),\nonumber
\end{eqnarray}
\begin{eqnarray}
\sqrt{2}A_{D^+\pi^0} &=& V_u ( f_\pi F^T_{e2} + M^T_{e2})-
V_c (f_{B_c} F^T_a + M^T_a) \nonumber\\
&&\hspace{-0.3cm}- V_t \left( f_\pi F^{P1}_{e2} +f_\pi F^{P2}_{e2}
+ f_\pi
F^{P3}_{e2}  \right.\nonumber\\
&& + M^{P1}_{e2}+ M^{P2}_{e2}+ M^{P3}_{e2} -f_{B_c} F^{P1}_a\nonumber\\
&&\left.  -f_{B_c} F^{P3}_a - M^{P1}_a - M^{P2}_a \right) ,
\label{dpi0}
\end{eqnarray}
where  $F({\cal M})$ denotes factorizable (non-factorizable)
amplitudes, the subscript $e(a)$ denotes the  emission
(annihilation) diagrams. The subscript $1(2)$ denotes the process
$B_c^+\to D^0\pi^+$ ($B_c^+\to D^+\pi^0$), the superscript $T(P)$
denotes amplitudes from the tree (penguin) operators, and
$f_{B_c}$ ($f_\pi$) is the ${B_c}$ ($\pi$) meson decay constant.
The detailed expressions of these amplitudes are shown in Appendix
\ref{apa}.

 From eq.(\ref{d0pi},\ref{dpi0}), we can see that unlike $B^\pm$,
 $B^0(\bar B^0)$ decays, we have three kinds of decay amplitudes
 with different weak and strong phases: penguin contributions proportional to
 $V_t$
 and two kinds of tree contributions proportional to $V_c$ and
 $V_u$, respectively. The interference between them gives   large
 direct CP violation which will be shown later.

As stated in the introduction, the two diagrams in Fig.1 give the
contribution for $B_c \to D $ transition form factor, which is
defined as
\begin{equation}
\langle D| d\gamma^\mu b| B_c\rangle = F_+
(p_{B_c}^\mu+p_D^\mu)+F_-(p_{B_c}^\mu-p_D^\mu) \;.
\end{equation}
We calculate $F_+$ in PQCD and get:
\begin{eqnarray}
F_+ &=& {4 f_B \over \sqrt{2 N_c}}\pi C_F M^2_{B_c}\int^1_0 dx_2
\int^\infty_0
b_2  d  b_2 \, \phi_D(x_2,b_2)  \nonumber\\
&& \times  \Big\{  \left[ 2 r_b -x_2  -( r_b-2 x_2 )r_2+(x_2 -2
r_b)r_2^2
\right] \nonumber\\
&&\hspace{1.2cm} \times \alpha_s(t_e^{(1)}) S_D(t_e^{(1)})
H_{e1}(\alpha_e,\beta_{e1},b_2)
\nonumber\\
&&\hspace{0.4cm} +  \left[(1-x_1 ) r_2 (2-r_2) \right]\nonumber\\
&&\hspace{1.2cm} \times \alpha_s(t_e^{(2)}) S_D(t_e^{(2)})
H_{e2}(\alpha_e,\beta_{e2},b_2) \Big\},\label{form}
\end{eqnarray}
which is the similar expression as $F_{e1}^T$ without Wilson
coefficients in the appendix. The numerical results of $F_+$ can
be found in Table~\ref{tab:form}: the form factor $F_+$ is
$0.169^{+0.05}_{-0.15} $ including the uncertainty of $\omega_D$,
which is comparable with previous calculations \cite{du,form}.

\begin{table}[thbp]
\begin{center}
\begin{tabular}[t]{c|ccc}
 \hline \hline
 $\omega_D$ & $0.40{\rm GeV}$ &$0.45{\rm GeV}$ &$0.50{\rm
  GeV}$ \\

\hline
$F_+$&  0.154 &  0.169& 0.174\\
\hline \hline
\end{tabular}
\end{center}
\caption{Form factor $F_+$ in the different values of $\omega_D$
.} \label{tab:form}
\end{table}

\subsection{Input parameters}

For $D^{}$ meson wave function,  two types of $D$ meson wave
function are usually used in the past literature:  one is
 \cite{li-bd}
\begin{eqnarray}
\phi_{D^{}}(x) &=&  \frac{3}{\sqrt{2 N_c}}
f_{D^{}} x(1-x)\{ 1 + a_{D^{}} (1 -2x) \}\nonumber\\
&&\times \exp
\left[-\frac{1}{2} (\omega_{D^{}} b)^2 \right],\label{wfd1}
\end{eqnarray}
in which the last term, $\exp \left[-\frac{1}{2}
(\omega_{D^{}}b)^2 \right]$, represents the $k_T$ distribution;
the other   \cite{li-lu,lu-song} is
\begin{eqnarray}
\phi_{D^{}}(x) &=&  \frac{3}{\sqrt{2 N_c}}
f_{D^{}} x(1-x)\{ 1 + a_{D^{}} (1 -2x) \},\label{wfd2}
\end{eqnarray}
which is fitted from the measured $B \to D^{}\ell \nu$ decay
spectrum at large recoil. The absence of the last term in the
Eq.(\ref{wfd1}) is due to the insufficiency of the experimental data.

\begin{figure}[t]
\vskip -1cm
   \epsfxsize=10cm
   \centerline{\epsffile{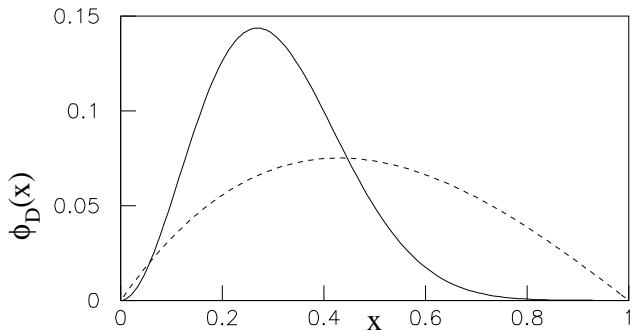}}
   \vspace*{-1.2cm}
\caption[dummy]{\label{wf}$D$ meson wave functions: the dashed line for
case 1 \&2, the solid line  for case 3.}
\end{figure}

Though the wave function of $D$ meson turns more complicated when
it runs at  a velocity of about $0.6c $, the light quark's
momentum must be less than $p^+_2/2$ because the mass of $c$ quark
is by far larger than $\Lambda_{\rm QCD}$: $M_c \gg \Lambda_{\rm
QCD}$, so the wave function of $D$ meson should be strongly
suppressed in the region $x_2>1/2$ even the $k_T$ contributions
are considered. In order to satisfy the above condition, we give
up the $D$ wave functions above and construct a new wave function,
which also fits the measured $B\to D l\nu$ decay spectrum at large
recoil:
\begin{eqnarray}
\phi_D(x,b) &=& N_D \, \left[ x(1-x)\right] ^2 \times\nonumber\\
&& \exp \left[ - {1\over 2} \left( {x M_D\over \omega_D} \right)
^2 - {1\over 2} \left( \omega_D\right) ^2 b^2 \right] ,
\end{eqnarray}
where $N_D$ is a normalization constant to let
\begin{eqnarray}
\int^1_0 \phi_D (x,b) &=& {f_D\over 2 \sqrt{2 N_c}} .
\end{eqnarray}
The behavior of all $D$ meson wave function can be seen in the
Fig.\ref{wf}. Our choice of the third case has a broad peak at the
small $x$ side, which characterize the mass difference of $m_c$
and $m_d$.

The $\pi$ wave functions   \cite{18ps,19vect} we adopt are
calculated by QCD sum rules and shown in the Appendix \ref{apb}.

The other input parameters are listed below \cite{chang,20pdg}:
\begin{eqnarray}
& & f_{B_c} = 480\; {\rm MeV},\;\; f_{D} = 240\; {\rm
MeV},\;\;
f_\pi = 131\; {\rm MeV},\;\; \nonumber\\
&&\omega_{D^{}}=0.45\;{\rm GeV},\;\;
M_{0\pi} = 1.60\; {\rm GeV},\;\;
a_{D^{}} = 0.3,\;\; \nonumber\\
&& M_{B_c} = 6.4\; {\rm GeV},\;\; M_b = 4.8\; {\rm GeV},\;\; \nonumber\\
&&M_D = 1.869\; {\rm GeV},\;\;
 M_t=170\; {\rm GeV},\;\;\nonumber\\
 &&M_W = 80.4\;{\rm GeV},\;\;\tau_{B^\pm}=0.46 \times 10^{-12}{\rm
 s},\nonumber\\
 &&G_F=1.16639\times 10^{-5}\;{\rm GeV}^{-2}. \label{eq:parapdg}
\end{eqnarray}

The CKM parameters used in the paper are:
\begin{eqnarray}
\left|{V_{ub}\over V_{cb}}\right| &=& 0.085 \pm  0.020,\label{ckm1} \\
|V_{cb}| &=& 0.039\pm 0.002, \\
R &=& \left|{V_u\over V_c}\right| = {1-\lambda^2/2 \over \lambda} \left|
{V_{ub}\over V_{cb}}\right|.
\end{eqnarray}
The CKM angle $\phi_3=\gamma$ is left as a free parameter to
discuss CP violation, defined by
\begin{eqnarray}
\gamma  &=& \arg \left(-{V_u\over V_c} \right)=\arg (V^\ast _{ub}) \label{ckm2}.
\end{eqnarray}

\subsection{Numerical analysis}

We fix  $\gamma=55^\circ$ to discuss the central value of
numerical results first.

Both process $B_c^+\to D^0\pi^+$ and $B_c^+\to D^+\pi^0$  are
tree-dominated. The branching ratios and main contributions are
give in Table \ref{num1}, from which we can see that the branching
ratio of $B_c^+\to D^0\pi^+$ is much larger than that of $B_c^+\to
D^+\pi^0$.  Though they  are both tree-dominated process, their
branching ratios and percentage of different topologies in the
whole process are obviously different. Because the  annihilation
topology give the same contributions to both processes, despite a
$\sqrt{2}$ factor, the difference only comes   from the emission
topology. In the process $B_c^+\to D^0\pi^+$, contributions from
factorizable emission topology dominate the whole tree
contributions for the large Wilson Coefficients $ C_2 +C_1/N_c$ in
Eq.(\ref{fet1}),   which occupy about $93\%$ of total even when
the effect of CKM matrix element is
considered.($|\lambda_u|<|\lambda_c|$). On the contrary,
contributions   from factorizable emission topology in the process
$B_c^+\to D^+\pi^0$ are suppressed because the Wilson Coefficients
$C_1 $ and $C_2/N_c$ in Eq.(\ref{fet2}) cancel each other
approximately. From the Table \ref{num1} we also find that
contributions
  from factorizable annihilation topology are at the same order of
   non-factorizable emission topology.

\begin{table}[thbp]
\begin{center}
\begin{tabular}[t]{c|c@{\qquad}c}
 \hline \hline
  $ $ & $B_c^+\to D^0\pi^+$ & $B_c^+\to D^+\pi^0$ \\
\hline
$f_\pi F^T_e  $&$23.0$&$0.763$\\
$M^T_e        $&$-0.379+0.863i$&$0.854-2.16i$\\
$f_B F^T_a    $&$-3.35+5.49i$&$-3.35+5.49i$\\
$M^T_a        $&$2.52-1.92i$&$2.52-1.92i$\\
\hline
$\displaystyle{\left|P\over T_e\right|}$&$10\%$&$40\%$\\
\hline
$Br$&$0.978\times 10^{-6}$&$0.196\times 10^{-6}$\\
\hline \hline
\end{tabular}
\end{center}
\caption{Branch ratios and main contributions   from tree
operators ($10^{-3}{\rm GeV}$ ).} \label{num1}
\end{table}

The ratio of the penguin contributions over the tree contributions
is about $10\%$ in the process $B_c^+\to D^0\pi^+$ and about $40\%$
in the process $B_c^+\to D^+\pi^0$ (Table \ref{num1}). The reason
for the difference is the following: the term $2 r_\pi
\phi^p_\pi(x_3)$ in the Eq.(\ref{fap3})   from $O_6,O_8$ operator
contributions, having no factors like $x_3$ to suppress its
integral value in the end-point region and leading to large
enhancement compared with other penguin contributions. But the
most important reason is that the tree contribution is suppressed
in the process $B_c^+\to D^+\pi^0$ due to small Wilson
coefficients $C_1+ C_2/3$ but not suppressed in the process
$B_c^+\to D^0\pi^+$. The $O_6,O_8$ contributions also affect the
dependence behavior of the branching ratio and the direct CP
asymmetry on the CKM angle $\gamma$ in the process $B_c^\pm\to
D^\pm\pi^0$, which will be discussed in the following.

\begin{figure}
\vskip -1.0cm
\begin{minipage}[t]{0.43\textwidth}
   \epsfxsize=8cm
   \centerline{\epsffile{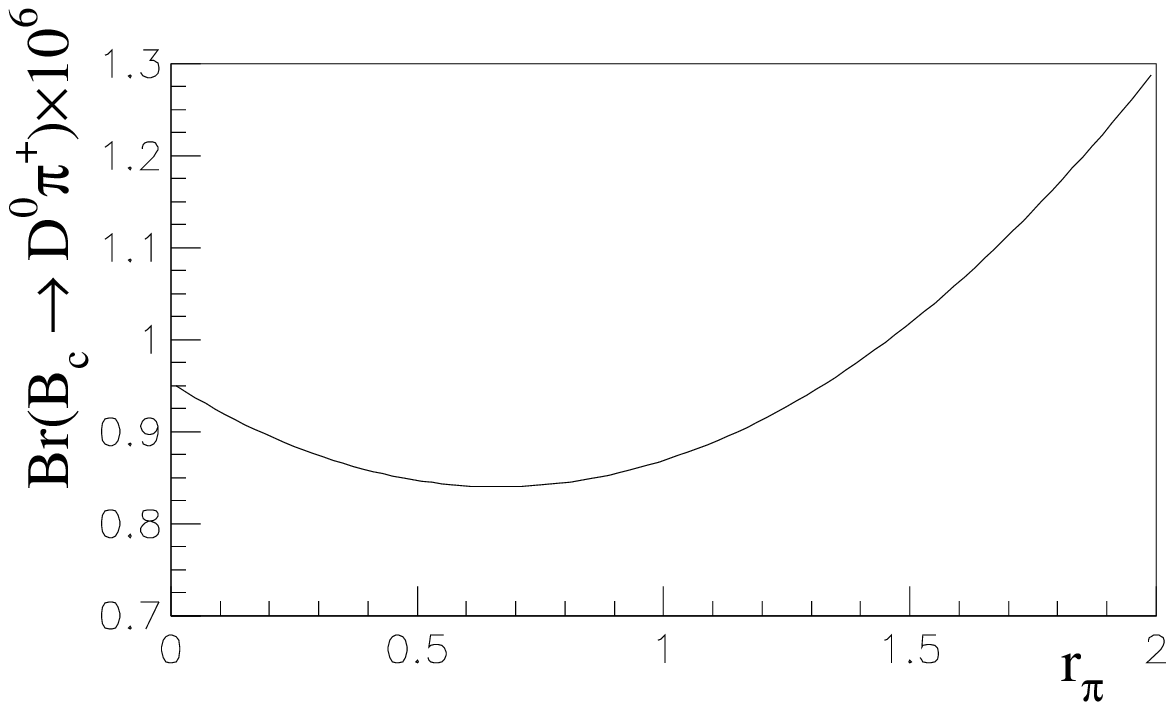}}
   \end{minipage}
\vskip -1.0cm
\begin{minipage}[t]{0.43\textwidth}
   \epsfxsize=8cm
   \centerline{\epsffile{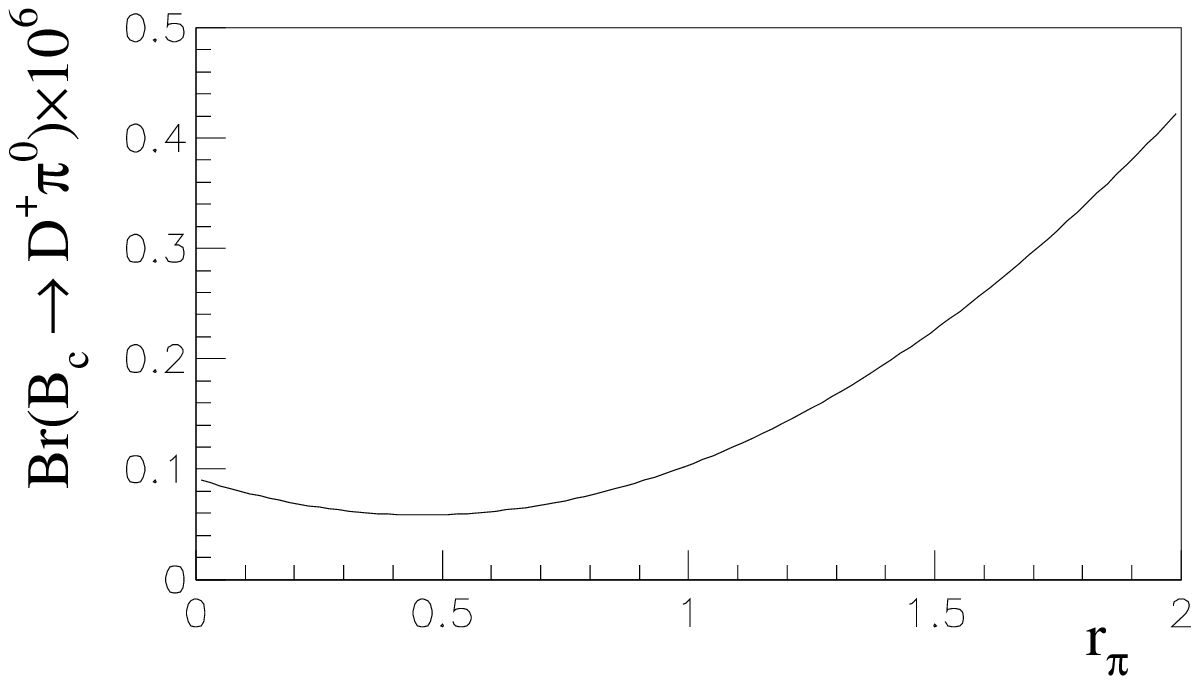}}
   \end{minipage}
   \vspace*{-0.8cm}
\caption[dummy]{\label{rpi}The correlation between $Br(B_c\to
D\pi)$ and $r_\pi$.}
\end{figure}

The correlation  between  $Br(B_c^+\to D\pi)$ and $r_\pi$ is shown
in Fig.\ref{rpi}. Because twist$-3$ terms of $\pi$ wave function
do not contribute to the form factor
(Eq.(\ref{form1},\ref{form2})), the variation of $r_\pi$ affects
the process  $B_c^+\to D^+\pi^0$ more heavily than the process
$B_c^+\to D^0\pi^+$, where the latter dominated by the $B_c \to D$
form factor diagrams. When $r_\pi=1.4$, the twist-3 contributions
are about $25\%$ in the process  $B_c^+\to D^0\pi^+$.
 In the process $B_c^+\to D^+\pi^0$, the twist-3 contributions with a relative
 minus sign cancel some of
the twist-2 contributions. When $r_\pi=1.4$, the branching ratio
of $B_c^+\to D^+\pi^0$ is about four times of the branching ratio
with only twist-2 contributions. When $r_\pi=0$, the twist-3
contributions vanish and only the contributions   from twist-2
terms in the $\pi$ wave function  are left. The corresponding
branching ratio is reduced to $0.95\times 10^{-6}$ in the process
$B_c^+\to D^0\pi^+$ and $0.092\times 10^{-6}$ in  the process
$B_c^+\to D^+\pi^0$ respectively.

\begin{table}[htbp]
\begin{center}
\begin{tabular}[t]{c|c@{\qquad}c}
 \hline \hline
  $ $ & $B_c^+\to D^0\pi^+$ & $B_c^+\to D^+\pi^0$ \\
\hline
$\omega_D=0.40{\rm GeV}  $&$ 1.03$   &$ 0.128$\\
$\omega_D=0.45{\rm GeV}  $&$  0.978$    &$ 0.196$\\
$\omega_D=0.50{\rm GeV}  $&$ 1.19$   &$ 0.199$\\
\hline \hline
\end{tabular}
\end{center}
\caption{Branch ratios in the unit $10^{-6}$  for
different $\omega_D$ .} \label{omega}
\end{table}

As the only free parameter with large uncertainty,  the value of
$\omega_D$ is the key point to  the whole prediction in the
calculations of $B_c\to D \pi $. In the Table \ref{omega} we
discuss the branching ratio in three groups of different $\omega$
values: $\omega_D=0.40{\rm GeV}$, $\omega_D=0.45{\rm GeV}$ and
$\omega_D=0.50{\rm GeV}$,   from which we see that the variation
of $\omega_D$ affects the process $B_c^+\to D^0\pi^+$ slightly,
but affect  the process $B_c^+\to D^+\pi^0$ heavily.

According to the CKM parametrization shown in the
Eq.(\ref{ckm1}-\ref{ckm2}), the decay amplitudes of $B_c\to D\pi$ can be
written as:
\begin{eqnarray}
M_{D\pi} &=& V_u T_u + V_c T_c - V_t P \nonumber\\
&=& V_u (T_u +P) \left[ 1-{1\over R}{T_c+P\over T_u+P}
e^{-i\gamma} \right] \nonumber\\
&\equiv& V_u (T_u+P) \left[ 1-z e^{i( -\gamma +\delta)}  \right] ,
\label{dp1}
\end{eqnarray}
where $z = {1\over R} \left| {T_c+P\over
T_u+P}\right|=\left|{V_c\over V_u}\right|\left| {T_c+P\over
T_u+P}\right|$   and the strong phase $\delta=$ $\arg\left(
T_c+P\over T_u+P\right) $,  from our PQCD calculation the
numerical value of which is $0.28$ and $123^\circ$ for  $ B_c^+\to
D^0\pi^+$, respectively. The emission topology in this channel is
only  about one time larger than the annihilation topology due to
the small CKM factor $|V_u/V_c|$.

\begin{figure}[htb]
\vskip -1cm
   \epsfxsize=10cm
   \centerline{\epsffile{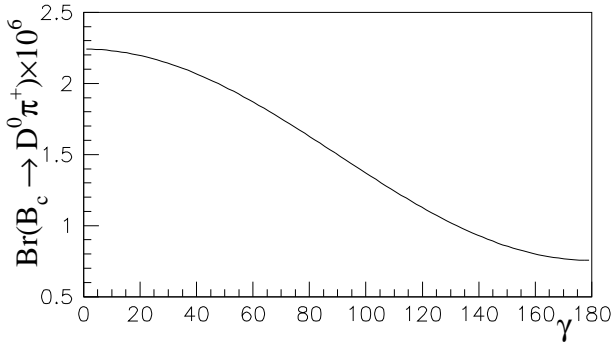}}
   \vspace*{-1.2cm}
\caption[dummy]{\label{br-va}The correlation between the averaged
branching ratio and $\gamma$ in the process $B_c^\pm\to
D^0\pi^\pm$. }
\end{figure}

The corresponding conjugate decay of $B_c^+\to D\pi$ reads:
\begin{eqnarray}
M_{ {  B}_c^-\to \bar D\pi^{-(0)}} &= & V^*_u (T_u+P) \left[ 1-z
e^{i( \gamma +\delta)}  \right] \, \label{conju}
\end{eqnarray}
and the averaged branching ratio for $B_c^\pm\to D^0 (\bar
D^0)\pi^\pm$ reads:
\begin{eqnarray}
Br  &=& {1\over 2} ( |M|^2 +{|\bar M|}^2 )\nonumber\\
&=&{1\over 2} |V_u (T_u+P) | ^2 \left[ 1 - 2 z \cos \gamma\cos
\delta +z^2 \right] ,\label{br}
\end{eqnarray}
which is the function of CKM angle $\gamma$. Its numerical result
depends on $\gamma$ significantly: the larger $\gamma$, the
smaller the averaged branching ratio, because $\cos\delta<0$. The
explicit correlation between the averaged branching ratio
$B_c^\pm\to D^0(\bar D^0)\pi^\pm$ and $\gamma$ is shown in the
Fig.\ref{br-va}.

\begin{figure}[htb]
\vskip -1cm
   \epsfxsize=10cm
   \centerline{\epsffile{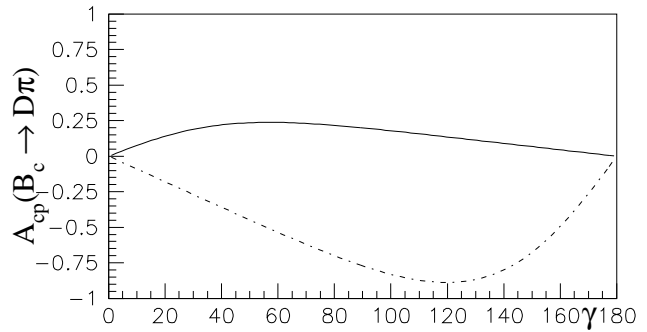}}
   \vspace*{-1.2cm}
\caption[dummy]{\label{fig:acp}The correlation between the direct
CP violation and $\gamma$, the solid line for $B_c^\pm\to
D^\pm\pi^0$ and the dashed line for $B_c^\pm\to D^0(\bar
D^0)\pi^\pm$. }
\end{figure}

The direct CP violation $A^{\rm dir}_{cp}$ is defined as
\begin{eqnarray}
A^{\rm dir}_{cp} = & {\left|M(B_c^+\to D^{0(+)}
\pi^{+(0)}\right|^2-\left|{ M}(B_c^-\to D^{0(-)}
\pi^{-(0)}\right|^2\over\left|M(B_c^+\to D^{0(+)}
\pi^{+(0)}\right|^2 +\left|{ M}(B_c^-\to D^{0(-)}
\pi^{-(0)}\right|^2}. \label{acp}
\end{eqnarray}
There are  two different tree contributions and one kind of
penguin contribution with different strong and weak phases, which
will contribute to the CP asymmetry. Using
eq.(\ref{dp1},\ref{conju}), $A^{\rm dir}_{cp}$  can be simplified
as
\begin{eqnarray}
A^{\rm dir}_{cp} &=&  -{2z\sin \delta \sin \gamma  \over 1 - 2 z
\cos\delta \cos \gamma + z^2 },\label{acp2}
\end{eqnarray}
which is proportional to $\sin\gamma$ approximately. This is shown
in Fig.\ref{fig:acp}. When $\gamma=55^\circ$, the direct CP
asymmetry is about $-50\%$ in the process $B_c^+\to D^0\pi^+$.

\begin{figure}[tb]
\vskip -1cm
   \epsfxsize=10cm
   \centerline{\epsffile{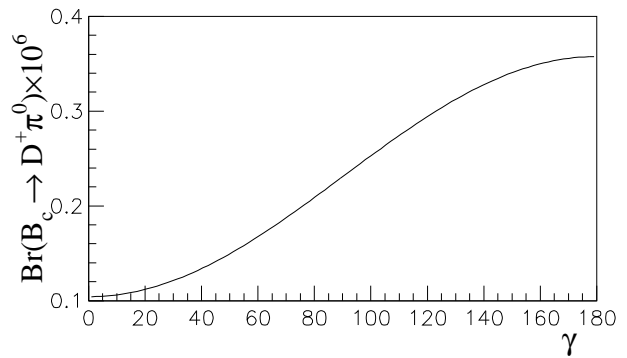}}
   \vspace*{-1.2cm}
\caption[dummy]{\label{br-va2}The correlation between the averaged
branching ratio and $\gamma$ in the process $B_c^\pm\to
D^\pm\pi^0$. }
\end{figure}

The $B_c\to D^+\pi^0$ becomes a little more complicated:  the tree
contributions   from the emission topology $M_e^T$ (in Table I) is
suppressed due to the small Wilson coefficients $C_1+C_2/3$.    In
this case, the three different contributions with different weak
and strong phases (two tree contributions and one penguin
contributions) are at the same order of magnitude.  We can still
use Eq.(\ref{br}) and Eq.(\ref{acp2}) to get the behavior of the
branching ratio and the direct CP asymmetry on $\gamma$. Now the
numerical values of $z$ and $\delta$ are $3.1$ and $-20^\circ$
respectively. Different from the averaged branching ratio of the
process $B_c\to D^+\pi^0$, the averaged branching ratio of the
process $B_c\to D^0\pi^+$ becomes smaller when $\gamma$ becomes
larger because $\cos\delta > 0$. The behavior of the branching
ratio and the direct CP asymmetry does not change much, but the
shape of the former turns more sharp. In one word, the branching ratio
 of $B_c^\pm\to D^\pm\pi^0$ shown in
Fig.\ref{br-va2} are more  sensitive to the value of $\gamma$,
which is quite different from the case for $B_c^\pm\to
D^0\pi^\pm$, but the direct CP asymmetry of  $B_c^\pm\to
D^\pm\pi^0$ shown in Fig.\ref{fig:acp} does not change greatly
because the large uncertainty from $\gamma$ cancels in the ratio
of  the direct CP asymmetry. When $\gamma=55^\circ$, the direct CP
asymmetry is about $25\%$ in the process $B_c^\pm\to D^\pm\pi^0$.
As pointed out in ref.\cite{next}, the CP asymmetry is sensitive
to the next-to-leading order contribution, which is more
complicated, the result shown here should be taken carefully.

\section{Conclusion}

In this paper we discuss the process $B_c\to D^0 \pi^+ $ and
$B_c\to D^+ \pi^0 $  in the PQCD approach and get their branching
ratios $1.03^{+0.16}_{-0.04}\times 10^{-6}$ and
$1.96^{+0.03}_{-0.68} \times 10^{-7}$ respectively. We also
predict the possible large direct CP asymmetry in the two
processes: $A^{\rm dir}_{cp}(B_c^\pm\to D^0 \pi^\pm )\approx
-50\%$ and $A^{\rm dir}_{cp}(B_c^\pm\to D^\pm \pi^0 )\approx 25\%$
when $\gamma=55^\circ$. The possible theoretical uncertainty are
also analyzed. We hope it can be tested in the coming experiments
at Tevatron, LHC and the super-B factory.

\section*{Acknowledgments}

One of the authors (JFC) thank C.H. Chang,  Y. Li, Y.-L. Shen and
X.-Q. Yu for the beneficial discussions. This work is partly
supported by National Science Foundation of China. \onecolumngrid

\appendix

\begin{appendix}

\section{Contributions   from all the diagrams}
\label{apa}

\subsection{Contributions   from factorizable diagrams}

All diagrams are sorted into two kinds: emission topology and
annihilation topology shown in Fig.\ref{fe},\ref{fa} and
Fig.\ref{nfe},\ref{nfa}. The factorizable tree contributions
  from emission topology read:
\begin{eqnarray}
F^{T(P1,P2)}_{ei} &=& {4 f_B \over \sqrt{2 N_c}}\pi C_F M^2_{B_c}\int^1_0 dx_2
\int^\infty_0
b_2  d  b_2 \, \phi_D(x_2,b_2)  \nonumber\\
&& \times  \Big\{  \left[ 2 r_b -x_2  -( r_b-2 x_2 )r_2+(x_2 -2 r_b)r_2^2
\right]
E_{ei}^{T(P1,P2)}(t_e^{(1)}) H_{e1}(\alpha_e,\beta_{e1},b_2)
\nonumber\\
&&+  \left[(1-x_1 ) r_2 (2-r_2) \right]E_{ei}^{T(P1,P2)}(t_e^{(2)})
H_{e2}(\alpha_e,\beta_{e2},b_2) \Big\},\label{form1}   \\
F^{P3}_{ei} &=& -{8 f_B \over\sqrt{2 N_c}} r_K\pi C_F M^2_{B_c}
\int^1_0  dx_2  \int^\infty_0
b_2  d  b_2\, \phi_D(x_2,b_2)  \nonumber\\
&& \times  \Big\{  \left[ -2 + r_b + ( 1-4 r_b+x_2 ) r_2 + (r_b-2 x_2 +2)
r_2 ^2 \right]
E_{ei}^{P3}(t_e^{(1)}) H_{e1}(\alpha_e,\beta_{e1},b_2)
\nonumber\\
&&-   \left[  x_1  + 2(1- 2 x_1) r_2-(2-x_1) r_2^2\right]E_{ei}^{P3}(t_e^{(2)})
 H_{e2}(\alpha_e,\beta_{e2},b_2)\Big\}.\label{form2}
\end{eqnarray}
Because $b$ and $c$ are both massive quarks, there is no collinear
divergence in the $B_c\to D$ transition, so the threshold resummation
needn't to be considered. In all the expressions, $T$ denotes the contributions
  from tree operators, $P1$ denotes the penguin contributions with the Dirac
structure $(V-A)\otimes (V-A)$, $P2$ denotes the penguin
contributions with the Dirac structure $(V-A)\otimes (V+A)$, and
$P3$ denotes the penguin contributions with the Dirac structure
$(S-P)\otimes (S+P)$; the subscript $e(a)$ denotes the
factorizable emission (annihilation) diagrams, the subscript $ne$
($na$) denotes the nonfactorizable emission (annihilation)
diagrams.

The factorizable tree contributions   from annihilation topology
read:
\begin{eqnarray}
F^{T(P1)}_{a} &&\hspace{-0.4cm} = 8 \pi C_F M^2_{B_c} \int^1_0  d x_2 d x_3
\int^\infty_0 b_2 d b_2 b_3 d b_3\, \phi_D(x_2 ,b_2)  \nonumber\\
&&\times \Big\{  \left[ \big( x_3 - (1+2 x_3 )r_2^2\big)
 \phi_\pi(x_3)+ r_2 r_\pi \big(
(1+2 x_3)\phi_\pi^p(x_3)   -(1-2 x_3)\phi_\pi^\sigma(x_3) \big) \right]
\nonumber\\
&&\hspace{1.2cm} \times E_{a}^{T(P1)}(t_a^{(1)})
H_a(\alpha_a,\beta_{a1},b_2,b_3) S_t(x_3)  \nonumber\\
&&- \left[ x_2 (1-r_2^2) \phi_\pi(x_3)+2 r_2 r_\pi (1+ x_2) \phi_\pi^p(x_3)
\right] E_{a}^{T(P1)} (t_a^{(2)}) H_a(\alpha_a,\beta_{a2},b_3,b_2)
S_t(x_2)  \Big\},  \\
F^{P3}_{a} &&\hspace{-0.4cm} = -16 \pi C_F M^2_{B_c} \int^1_0  d x_2 d x_3
\int^\infty_0 b_2 d b_2 b_3 d b_3 \,\phi_D(x_2 ,b_2)  \nonumber\\
&&\times \Big\{  \left[  - r_2 \phi_\pi(x_3) +r_\pi \big(-x_3+ ( 2+ x_3 ) r_2^2
\big) \phi_\pi^p + r_\pi x_3 (1-r_2^2) \phi_\pi^\sigma(x_3) \right] \nonumber\\
&&\hspace{1.2cm} \times E^{P3}_{a} (t^{(1)}_a)
H_a(\alpha_a,\beta_{a2},b_2,b_3)S_t(x_3)                       \nonumber\\
&&- \left[ x_2r_2  \phi_\pi(x_3) + 2 r_\pi \big( 1- ( 1-x_2 )r_2^2 \big)
\phi_\pi^p(x_3) \right] E_{a}^{P3} (t_a^{(2)})
H_a(\alpha_a,\beta_{a2},b_3,b_2) S_t(x_2) \Big\}, \label{fap3}
\end{eqnarray}
where the factor $S_t(x)$ is the jet function   from the threshold
resummation  \cite{thre}:
\begin{equation}
 S_t(x) = \frac{2^{1+2c}\Gamma(3/2 +c)}{\sqrt{\pi} \Gamma(1+c)}
[x(1-x)]^c,\quad c = 0.3.
\end{equation}
The factors $E_i^{T(P)}(t)$ contain the Wilson coefficients $a(t)$
at scale $t$ and the evolution   from $t$ to the factorization
scale $1/b$ in the Sudakov factors $S(t)$:
\begin{eqnarray}
E_{ej}^{T(Pi)} (t) &=& \alpha_s (t) a_{ej}^{T(Pi)} (t)\,  S_D(t), \nonumber\\
E_{a}^{T(Pi)} (t) &=& \alpha_s (t) a_{e1}^{T(Pi)} (t) \, S_{D} (t)
S_\pi(t),
\end{eqnarray}
where $S_D(t), S_\pi(t)$, the Sudakov factors, are defined as
\begin{eqnarray}
& & S_D(t) = s(x_2P_2^+,b_2) +
2 \int_{1/b_2}^t \frac{d\mu}{\mu} \gamma_q(\mu), \\
& &S_\pi(t) = s(x_3P_3^-,b_3) + s((1-x_3)P_3^-,b_3) + 2
\int_{1/b_3}^t \frac{d\mu}{\mu} \gamma_q(\mu),
\end{eqnarray}
and $s(Q,b)$ is given as \cite{23lm}
\begin{eqnarray}
  s(Q,b) &=& \int_{1/b}^Q \!\! \frac{d\mu}{\mu} \left[
 \left\{ \frac{2}{3}(2 \gamma_E - 1 - \ln 2) + C_F \ln \frac{Q}{\mu}
 \right\} \frac{\alpha_s(\mu)}{\pi} \right. \nonumber \\
& &  \left.+ \left\{ \frac{67}{9} - \frac{\pi^2}{3} -
\frac{10}{27} n_f
 + \frac{2}{3} \beta_0 \ln \frac{e^{\gamma_E}}{2} \right\}
 \left( \frac{\alpha_s(\mu)}{\pi} \right)^2 \ln \frac{Q}{\mu}
 \right],
 \label{eq:SudakovExpress}
\end{eqnarray}
where the Euler constant $\gamma_E=0.57722\cdots$, and $\gamma_q =
-\alpha_s/\pi$ is the quark anomalous dimension.

 The hard functions $H$ are
\begin{eqnarray}
H_{e1}(\alpha,\beta,b) &=& {K_0(\alpha b)- K_0(\beta
b)\over \beta^2-\alpha^2 },\\
H_{e2}(\alpha,\beta,b) &=&{1\over (1-x_1 )(x_1 -r_2^2)} {K_0(\alpha b)},\\
H_a( \alpha,\beta,b_1,b_2) &=& \left[ \theta(b_1-b_2) K_0(\alpha b_1)
I_0(\alpha  b_2)+  \theta(b_2-b_1) K_0(\alpha b_2)
I_0(\alpha  b_1)\right] K_0(\beta b_2 ),
\end{eqnarray}
where $K_0, I_0, H_0$ and $J_0$ are the Bessel functions of order 0. It is
implied that the
transformed Bessel functions $K_0$ and $I_0$  become the corresponding
  Bessel functions with real variable  when their variables are complex.

The Wilson Coefficients $a_i$ read:
\begin{eqnarray}
a^T_{e1} (t) &=& C_2+ {C_1\over N_c},\label{fet1}\\
a^{P1}_{e1} (t)&=& C_4+ {C_3\over N_c}  +C_{10}+ {C_9\over N_c}, \nonumber\\
a^{P3}_{e1}(t) &=&\left( C_6+ {C_5\over N_c}\right) + \left( C_8 +
{C_7\over N_c}
\right), \nonumber\\
a^T_{e2} (t)&=& C_1+ {C_2\over N_c},\label{fet2}\\
a^{P1}_{e2}(t) &=&-\left( C_4+{C_3\over N_c} \right) +{3\over
2}\left( C_9+{C_{10}\over N_c }
\right) + {1\over 2} \left( C_{10}+{C_9\over N_c}\right)  , \nonumber\\
a^{P2}_{e2}(t) &=&
-
 {3\over 2} \left( C_7 +{C_8\over N_c}\right), \nonumber\\
a^{P3}_{e2} (t)&=& -  \left( C_6+ {C_5\over N_c} \right)   +
{1\over 2} \left( C_8 + {C_7\over N_c} \right) .
\end{eqnarray}
All the Wilson coefficients $C_i$ above should be evaluated at the
appropriate scale $t$. The hard scale $t$'s are chosen as the
maximum of the virtuality of internal momentum transition in the
hard amplitudes, including $1/b_i$:
\begin{eqnarray}
t_e^{(1)} &=& \max \left(|\alpha_e|,|\beta_{e1}|,1/b_2 \right), \nonumber\\
t_e^{(2)} &=& \max \left(|\alpha_e|,|\beta_{e2}|,1/b_2 \right), \nonumber\\
t_a^{(1)} &=& \max \left(|\beta _{a1}|,1/b_2,1/b_3 \right),\nonumber\\
t_a^{(1)} &=& \max \left(|\beta_{a2}|,1/b_2,1/b_3 \right),\nonumber
\end{eqnarray}
where
\begin{eqnarray}
\alpha^2_e &=& (1-x_1 -x_2 )(x_1 -r_2^2)M^2_{B_c},\nonumber\\
\beta^2_{e1}&=& [r_b^2-x_2  (1-r^2_2)  ] M^2_{B_c},\nonumber\\
\beta^2_{e2}&=& (1-x_1 )(x_1 -r_2^2) M^2_{B_c},\nonumber\\
\alpha^2_a &=& -x_2  x_3 M^2_{B_c} (1-r^2_2), \nonumber\\
\beta^2_{a1} &=&- x_3  M^2_{B_c} (1-r^2_2), \nonumber\\
\beta^2_{a2}&=&- x_2  M^2_{B_c} (1-r^2_2). \nonumber\\
\end{eqnarray}

\subsection{Contributions   from non-factorizable diagrams}

Different   from factorizable diagrams, non-factorizable diagrams
include convolution of all three wave functions and, of course,
the convolution of  Sudakov factors. Their amplitudes are:
\begin{eqnarray}
M^{T(P1)}_{ei} &=&   {8\over N_c}\pi C_F f_B  M^2_{B_c} \int^1_0
d x_2  d x_3 \int^\infty_0  b_2 d b_2 b_3 d b_3  \phi_D(x_2,b_2)
\phi_\pi(x_3 ) \nonumber\\
&&\times \Big\{ \left[ 1- x_1 - x_3- (1-x_1 -x_2 )r_2 - (x_2 -2
x_3   ) r_2^2 \right] E^{T(P1)}_{ne~i}(t^{(1)}_a)\,
H_a(\alpha_{ne},\beta_{ne1},b_2,b_3)
\nonumber\\
&&+ \left[ (2 x_1  +x_2 -x_3 -1) + (1-x_1 -x_2 )r_2 +(-2 x_1 -x_2
+2 x_3 ) r_2^2 \right]E^{T(P1)}_{ne~i}(t^{(2)}_a)\,
H_a(\alpha_{ne},\beta_{ne2},b_2,b_3)
\Big\},\nonumber\\  \\
M^{P2}_{ei} &=&   {8\over N_c}\pi r_\pi C_F f_B  M^2_{B_c} \int^1_0
d x_2  d x_3 \int^\infty_0  b_2 d b_2 b_3 d b_3  \phi_D(x_2,b_2) \nonumber\\
&&\times \Big\{ \left[ \big( 1- x_1-x_3+(2-2 x_1 -x_2 -x_3 )r_2 +(1-x_1 -x_2
+x_3  )r_2^2 \big) \phi_\pi^p(x_3) \right.\nonumber\\
 &&\hspace{0.5cm}  +\left. \big( 1-x_1-x_3+(x_2 -x_3 ) r_2 +(-1+x_1 +x_2
 +x_3  )r_2^2  \big)\phi_\pi^\sigma(x_3)\right]  E^{P2}_{ne~i}(t^{(1)}_a)\,
 H_a(\alpha_{ne},\beta_{ne1},b_2,b_3)\nonumber\\
&&+ \left[ \big( x_1 - x_3 + (2 x_1  +x_2 -x_3 -1)r_2 +( x_1 +x_2
+x_3 -2)
r_2^2 \big) \phi_\pi^p(x_3) \right. \nonumber\\
&&\hspace{0.5cm} \left. + \big( -x_1 + x_3 + (x_2 +x_3 -1)r_2+(x_1
+x_2 -x_3 ) r_2^2 \big) \phi_\pi^\sigma(x_3) \right]
E^{P2}_{ne~i}(t^{(2)}_a)\,
H_a(\alpha_{ne},\beta_{ne2},b_2,b_3) \Big\}, \\
M^{P3}_{ei} &=&  {8\over N_c}\pi C_F f_B  M^2_{B_c} \int^1_0
d x_2  d x_3 \int^\infty_0  b_2 d b_2 b_3 d b_3  \phi_D(x_2,b_2)
\phi_\pi(x_3 ) \nonumber\\
&&\times \Big\{ \left[ -2+2 x_1 +x_2 +x_3 +(1-x_1 -x_2 )r_2+(2-2
x_1 -x_2 -2 x_3  )r_2^2 \right] E^{P3}_{ne~i}(t^{(1)}_a)\,
H_a(\alpha_{ne},\beta_{ne1},b_2,b_3)\nonumber\\
&&+ \left[ -x_1 +x_3 +(x_1 +x_2 -1)r_2-(x_2 +2 x_3 -2)  r_2^2
\right]E^{P3}_{ne~i}(t^{(2)}_a)\, H_a(\alpha_{ne},\beta_{ne2},b_2,b_3) \Big\},
\\
M^{T(P1)}_{a}&=&     {8\over N_c}\pi C_F f_B  M^2_{B_c} \int^1_0 d
x_2  d x_3 \int^\infty_0  b_2 d b_2 b_3 d b_3  \phi_D(x_2,b_2)
 \nonumber\\
&&\times \Big\{ \left[ (-x_1+x_2 + r_2 ) \phi_\pi(x_3)+ ( -2 x_1 + x_2 + x_3 + 4
r_2) r_2 r_\pi \phi_\pi^p(x_3)+ (x_2-x_3) r_2 r_\pi \phi_\pi^\sigma(x_3)\right]
\nonumber\\
&&\hspace{1.2cm} \times E^{T(P1)}_{ne1}(t^{(1)}_{na})\,
H_{na}(\alpha_{na},\beta_{na1},b_2) \nonumber\\
&&\hspace{0.5cm}+\big[ \big( 1-x_1-x_3 - r_b + ( -x_2 + 2 x_3 + r_b)r_2^2 \big)
\phi_\pi(x_3) + ( 2- 2 x_1 - x_2-x_3 - 4 r_b) r_2 r_\pi \phi_\pi^p(x_3)
\nonumber\\
&&\hspace{1.2cm} + ( x_2-x_3) r_2 r_\pi \phi_\pi^\sigma(x_3) \big]
E^{T(P1)}_{ne1}(t^{(2)}_{na})\, H_{na}(\alpha_{na},\beta_{na2},b_2) \Big\},\\
M^{P2}_{a}&=&- {8\over N_c}\pi C_F f_B  M^2_{B_c} \int^1_0 d x_2
d x_3 \int^\infty_0  b_2 d b_2 b_3 d b_3  \phi_D(x_2,b_2)
 \nonumber\\
&&\times \Big\{ \big[ (-x_1 + x_2-r_2) r_2 \, \phi_\pi(x_3) + \big(  x_1-x_3
+ r_2 + ( x_1-x_2 + x_3) r_2^2 \big) r_\pi  \phi_\pi^p(x_3) \nonumber\\
&&\hspace{0.5cm} +  \big( x_1+ x_2 - x_3 + ( -x_1+ x_2 + x_3)r_2^2
\big) r_\pi \,  \phi_\pi^\sigma(x_3) \big]
E^{P2}_{ne1}(t^{(1)}_{na})\,
H_{na}(\alpha_{na},\beta_{na1},b_2)\nonumber\\
&&- \big[ (-1 - r_b + x_1 +x_2 )r_2 \phi_\pi(x_3) + \big( 1+r_b -x_1
-x_3+(1+r_b - x_1-x_2 +x_3  ) r_2^2 \big) r_\pi \, \phi_\pi^p(x_3) \nonumber\\
&&\hspace{0.5cm}   + \big( 1+ r_b-x_1-x_3+ (-1 - r_b +x_1 +x_2 +
x_3 )r_2^2 \big) r_\pi \phi_\pi^\sigma(x_3) \big]
E^{P2}_{ne1}(t^{(2)}_{na})\,H_{na}(\alpha_{na},\beta_{na2},b_2)
\Big\}.
\end{eqnarray}
where the hard kernel $H_{na}$ is defined as
\begin{eqnarray}
H_{na}(\alpha,\beta,b) &=& {K_0(\alpha b)- K_0(\beta
b)\over \beta^2-\alpha^2 },
\end{eqnarray}
and the factor $E(t)$ turns into:
\begin{eqnarray}
E_{nej}^{T(Pi)} (t) &=& \alpha_s (t) a_{nej}^{T(Pi)} (t)\,
 S_{D}(t) S_\pi(t),
\end{eqnarray}
where  the Wilson coefficients $a$ read:
\begin{eqnarray}
a^T_{ne1}(t) &=&C_1 , \nonumber\\
a^{P1}_{ne1}(t) &=& C_3+ C_9, \nonumber\\
a^{P2}_{ne1}(t)&=& C_5+ C_7,\nonumber\\
a^T_{ne2} (t)&=& C_2, \nonumber\\
a^{P1}_{ne2}(t)&=& - C_3 + {1\over 2}C_9 + {3\over 2} C_{10} , \nonumber\\
a^{P2}_{ne2}(t)&=& -  C_5
+ {1\over 2}  C_7, \nonumber\\
a^{P3}_{ne2}(t)&=&{3\over 2}  C_8 .
\end{eqnarray}
The hard scale $t$'s are chosen as the maximum of the virtuality
of internal momentum transition in the hard amplitudes, including
$1/b_i$:
\begin{eqnarray}
t_e^{(1)} &=& \max\left(|\alpha_{ne}|,|\beta_{ne1}|,1/b_2,1/b_3\right), \nonumber\\
t_e^{(2)} &=& \max\left(|\alpha_{ne}|,|\beta_{ne2}|,1/b_2,1/b_3\right),\nonumber\\
t_a^{(1)} &=& \max \left(|\alpha_{na}|,|\beta _{na1}|,1/b_2 \right),\nonumber\\
t_a^{(1)} &=& \max \left(|\alpha_{na}|,|\beta_{na2}|,1/b_2 \right),\nonumber
\end{eqnarray}
where
\begin{eqnarray}
\alpha^2_e &=& (1-x_1 -x_2 )(x_1 -r_2^2)M^2_{B_c},\nonumber\\
\beta_{ne1}&=& - (1-x_1 -x_2 )\left[ (1-x_3 )(1-r^2_2)  -x_1 +r_2^2 \right]
M^2_{B_c},\nonumber\\
\beta_{ne2}&=& - (1-x_1 -x_2 )\left[ x_3  (1-r^2_2)  -x_1 +r_2^2 \right]
M^2_{B_c},\nonumber\\
\alpha^2_a &=& -x_2  x_3 M^2_{B_c} (1-r^2_2), \nonumber\\
\beta_{na1} &=& x_1  \left[ x_2 +x_3  (1-r^2_2)  \right] M^2_{B_c},
\nonumber\\
\beta_{na1}&=& (1-x_1 )\left[ x_2  + x_3  (1-r^2_2)  \right] M^2_{B_c}.
\end{eqnarray}

\section{The $\pi$ meson wave functions}
\label{apb}

 The different distribution amplitudes of $\pi$ meson wave functions are given as
 \cite{18ps,19vect}
 \begin{eqnarray}
 \phi_\pi(x) &=&  \frac{3}{\sqrt{6} }
  f_\pi  x (1-x)  \left[1+0.44C_2^{3/2} (2x-1) +0.25 C_4^{3/2}
  (2x-1)\right],\label{piw1}\\
 \phi_{\pi}^p(x) &=&   \frac{f_\pi}{2\sqrt{6} }
   \left[ 1+0.43 C_2^{1/2} (2x-1) +0.09 C_4^{1/2} (2x-1) \right]  ,\\
 \phi_{\pi}^\sigma(x) &=&  \frac{f_\pi}{2\sqrt{6} } (1-2x)
   \left[ 1+0.55  (10x^2-10x+1)  \right]  .    \label{piw}
 \end{eqnarray}

with the Gegenbauer polynomials
 \begin{equation}
 \begin{array}{l@{\qquad}l}
 \vspace{0.2cm}
 \displaystyle{C_2^{1/2} (t) = \frac{1}{2} (3t^2-1)}, & \displaystyle{C_4^{1/2} (t) = \frac{1}{8}
 (35t^4-30t^2+3)},\\
 \displaystyle{C_2^{3/2} (t) = \frac{3}{2} (5t^2-1)}, & \displaystyle{C_4^{3/2} (t) = \frac{15}{8}
 (21t^4-14t^2+1)}.
 \end{array}
 \end{equation}
\end{appendix}
\twocolumngrid


\begin{thebibliography}{99}
\bibitem{tevatron}CDF collaboration, F. Abe, {\it et al.},
Phys.\ Rev.\ D {\bf 58}, 112004 (1998).
\bibitem{pqcd}
Y.~Y.~Keum, H.~N.~Li and A.~I.~Sanda, Phys.\ Rev.\ D {\bf 63},
054008 (2001); C.~H.~Chen, Y.~Y.~Keum and H.~n.~Li, Phys.\ Rev.\ D
{\bf 64}, 112002 (2001); Y.~Y.~Charng and H.~n.~Li,
arXiv:hep-ph/0308257.

\bibitem{pqcd-lu} C.-D. L\"u, K. Ukai and M.-Z. Yang,
Phys. Rev. D63, 074009 (2001); C.-D. L\"u,   pp.  173-184,
Proceedings of International Conference on Flavor Physics (ICFP
2001), World Scientific, 2001, hep-ph/0110327.

\bibitem{qcdf}
M.~Beneke, G.~Buchalla, M.~Neubert and C.~T.~Sachrajda, Phys.\
Rev.\ Lett.\  {\bf 83}, 1914 (1999); Nucl.\ Phys.\ B {\bf 606},
245 (2001); M.~Beneke and M.~Neubert,
Nucl.\ Phys.\ B {\bf 675}, 333 (2003).

\bibitem{scet} C.W. Bauer, D. Pirjol, I.W. Stewart,  Phys. Rev. Lett. 87,
 201806 (2001); Phys. Rev. D65, 054022 (2002);
C.W. Bauer, I.W. Stewart, Phys. Lett. B516, 134 (2001);
     Phys.\ Lett.\ B {\bf 516}, 134 (2001); Phys.\ Rev.\
D {\bf 65}, 054022 (2002).

\bibitem{du}D. Du and Z. Wang,
Phys.\ Rev.\ D {\bf 39}, 1342 (1989);
\bibitem{nf}
J.Schwinger, Phys. Rev. Lett. {\bf 12}, 630 (1965);
M.Bauer and B. Stech, Phys. Lett. {\bf B152}, 380 (1985);
M.Bauer, B. Stech and M. Wirbel, Z. Phys. {\bf C34}, 103 (1987).
\bibitem{ali}
A. Ali and C. Greub, Phys. Rev. {\bf D57}, 2996 (1998); A. Ali,
J.Chay, C. Greub and P. Ko, Phys. Lett. {\bf B424}, 161 (1998); A.
Ali, G. Kramer and C. D. L$\ddot{\rm u}$, Phys. Rev. {\bf D58},
094009 (1998); Y.-H. Chen, H.-Y. Cheng, B. Tseng, K.-C. Yang,
Phys. Rev. D60  094014 (1999).

\bibitem{li-bd} T. Kurimoto, H.-n. Li, and A. I. Sanda,
Phys. Rev. D {\bf 67}, 054028 (2003).

\bibitem{kurimoto} C.D. L\"u and M.Z. Yang, Eur. Phys. J. C28, 515
(2003).
\bibitem{buras}
G.~Buchalla,  A.J.~Buras and M.E.~Lautenbacher, Rev.\ Mod.\ Phys.\
{\bf 68}, 1125 (1996).

\bibitem{form}J.F. Liu and K.T. Chao, Phys. Rev. D56, 4133
(1997); C.Q. Geng, C.W. Hwang and C.C. Liu, Phys. Rev. D65, 094037
(2002).
\bibitem{li-lu} Y.-Y. Keum, T. Kurimoto, H.-N. Li, C.-D. L\"u, A.I. Sanda,
Phys. Rev. D {\bf 69}, 094018 (2004).
\bibitem{lu-song} G.-L. Song and C.-D. L\"u,
Phys. Rev. D {\bf 70}, 034006 (2004).

\bibitem{18ps} P. Ball, JHEP, 09, 005, (1998); JHEP, 01, 010, (1999).


\bibitem{19vect} P. Ball, V.M. Braun, Y. Koike, and K. Tanaka, Nucl. Phys.
B 529, 323 (1998); P. Ball, V. M. Braun, hep-ph/9808229.
\bibitem{chang}C.-H. Chang and Y.Q. Chen, Phys. Rev. D 49, 3399 (1994).

\bibitem{20pdg} Particle Data Group, Phys. Rev. D66, Part I (2002).
\bibitem{next}  H.-n. Li, S. Mishima, A.I. Sanda, hep-ph/0508041.
\bibitem{thre} H.n. Li, Phys. Rev. D66, 094010 (2002).
\bibitem{23lm} H.-n. Li, B. Melic, Eur. Phys. J. C11, 695
(1999).
\end{thebibliography}
\end{document}